\documentclass[prd, twocolumn, nofootinbib, floatfix]{revtex4-1}

\usepackage{amsmath}
\usepackage{graphicx}
\usepackage{dcolumn}
\usepackage{bm}
\usepackage{epsfig}
\usepackage{amssymb,latexsym,mathrsfs}
\usepackage{graphicx}
\usepackage{color}
\usepackage{hyperref}

\hypersetup{
    colorlinks=true,
    linkcolor=red,
    citecolor=blue,
} 

\newcommand{\be}{\begin{equation}}
\newcommand{\ee}{\end{equation}}
\newcommand{\bs}{\begin{split}} 
\newcommand{\bea}{\begin{eqnarray}}
\newcommand{\eea}{\end{eqnarray}}

\newcommand{\om}{\Omega_m}
\newcommand{\op}{\Omega_{\phi}}

\newcommand{\finf}{f_\infty} 
\newcommand{\fcl}{F_{\rm cl}} 
\newcommand{\fcli}{F_{{\rm cl},\infty}} 
\newcommand{\fclt}{F_{{\rm cl},0}} 
\newcommand{\gstar}{g_\star} 
\newcommand{\sige}{\sigma_8} 
\newcommand{\siget}{\sigma_{8,0}}

\begin{document}

\title{Testing Dark Matter Clustering with Redshift Space Distortions} 
\author{Eric V.\ Linder} 
\affiliation{Berkeley Center for Cosmological Physics \& Berkeley Lab, 
University of California, Berkeley, CA 94720, USA\\ 
Institute for the Early Universe WCU, Ewha Womans University, 
Seoul 120-750, Korea}

\begin{abstract}
The growth rate of large scale structure can probe whether dark matter 
clusters at gravitational strength or deviates from this, e.g.\ due to 
self interactions.  
Measurement of the growth rate through redshift space distortions in 
galaxy redshift surveys constrains the clustering strength, and its 
redshift dependence.  We compare such effects on growth to those from 
high redshift deviations (e.g.\ early dark energy) or modified gravity, and 
give a simple, highly accurate analytic prescription.  Current observations 
can constrain the dark matter clustering strength to $F_{\rm cl}=0.99\pm0.02$ 
of standard, if all other parameters are held fixed, but substantial 
covariances exist.  Future galaxy redshift surveys may constrain an evolving 
clustering strength to 28\%, marginalizing over the other parameters, or 
4\% if the dark energy parameters are held fixed while fitting for dark 
matter growth.  
Tighter constraints on the nature of dark matter could be obtained by 
combining cosmological and astrophysical probes. 
\end{abstract}

\date{\today} 

\maketitle

%%%%%%%%%%%%%%%%%%%%%%%%%%%%%%%%%%%%%%%%%%%%%%%%%%%%%%%%%%%%%%%%%%%%%%%%
\section{Introduction} 

Growth of large scale structure in the universe carries rich cosmological 
information.  This includes the contributions of different energy density 
components, their evolution, and the strength of gravity and any other 
interactions.  With the increasing volume of galaxy surveys, probes 
of the cosmic growth history are being extended in redshift and precision. 
In particular, spectroscopic surveys enable measurement of redshift space 
distortion effects, proportional to the growth rate. 

Redshift space distortions were proposed as a test of the matter 
(and cosmological constant $\Lambda$) density in the 1980s and 1990s 
(see, e.g., \cite{kaiser,matsubara,ballinger,hamilton}), 
and then extended to reveal dark energy characteristics 
\cite{matsza,coorayhut}, and test modifications 
of gravity \cite{lin0709,guzzo}.  Here we investigate their use for 
exploring the nature of dark matter, specifically their clustering 
strength and the possible presence of self interactions.  

Since the physics of dark matter is unknown, we should not assume that 
it behaves identically to normal (baryonic) matter, as far as its cosmic 
evolution or interactions.  Numerous articles, e.g.\ 
\cite{mueller,dm1,dm2,dm3,dm4,dm5,dm6} 
have investigated the scaling 
of its energy density $\rho$ with expansion factor $a$ to test whether it 
follows the normal $\rho\sim a^{-3}$ evolution of pressureless matter, or 
has a nonzero equation of state (pressure to density) ratio. 

Others have explored nongravitational ``fifth'' forces, due to interactions 
between the dark matter and another sector of particle physics. 
One category would be a coupling between dark matter and dark energy 
(e.g.\ \cite{amen9908023} and many others).  
This adds issues of how this coupling originates (between components with 
vastly different energy scales), just what is the functional form of the 
coupling, how it is preserved under quantum corrections, and how and where 
dark matter is created or destroyed through the interaction (i.e.\ its 
velocity and density distribution relative to dark energy).  

Here we consider a simpler situation of exploring dark matter without any 
interaction with dark energy, but with possible self interaction 
that alters its clustering strength, while leaving its background evolution 
unchanged.  The analysis is intended to be a purely phenomenological 
assessment of what growth data can say about the clustering strength, but 
related effects can arise in models such as 
self interacting dark matter, e.g.\ 
\cite{gradwohl,spergelsteinh,0002050,12055809,12083026} 
multiple dark matter, e.g.\ 
\cite{0307316,07071495,07092297,10021306,12040514,13022614}, 
atomic dark matter \cite{12095752}, resonant dark matter \cite{12100900}, 
cannibal dark matter \cite{hall}, etc. 

Section \ref{sec:grow} investigates the effect of a deviation in the dark 
matter clustering strength 
on the density perturbation growth equations.  The results are compared 
to the analytic growth deviation formalism of the gravitational growth 
index $\gamma$ \cite{lingam} and growth calibration $g_\star$ \cite{gstar}, 
and the interacting growth formalism \cite{ameneta}, in Sec.~\ref{sec:anly}. 
We calculate the impact on the growth rate and redshift space distortions 
in Sec.~\ref{sec:data}, and use current and projected data to estimate 
constraints on the dark matter evolution.  Section~\ref{sec:concl} concludes.

%%%%%%%%%%%%%%%%%%%%%%%%%%%%%%%%%%%%%%%%%%%%%%%%%%%%%%%%%%%%%%%%% 
\section{Dark Matter Clustering and Growth} \label{sec:grow} 

The growth of density perturbations is driven by a source term of 
the gravitational potential inhomogeneities, and damped by 
the cosmic expansion.  Therefore the physical ingredients that affect 
growth are the expansion rate, or Hubble parameter $H=\dot a/a$, the 
inhomogeneities in energy density that source the Poisson equation, and 
any modification to gravitational attraction.  We take the dark energy 
component to remain effectively unclustered, as for standard scalar fields, 
so the only perturbation source is the matter density perturbations 
themselves.  

In the standard case, where all matter clusters with gravitational 
strength, the 
density perturbation source term is determined by $G_N\rho_m\delta$, where 
the density contrast $\delta=\delta\rho_m/\rho_m$, $G_N$ is Newton's 
constant, and $\rho_m$ is the total matter density.  We will allow the 
clustering to be altered, such that the source term is $G_N\fcl\rho_m\delta$. 
This can be interpreted as only a fraction $\fcl$ participating in 
clustering, or a clustering attraction varying from gravitational strength, 
or a combination of the two; we generically call $\fcl$ the clustering 
strength.  We do not intend it to be viewed as a modified 
gravity model, where $G_N\to G_N\fcl$ since this would in general need to be 
formed from an action principle to establish the new equations of motion. 
Thus we stay within general relativity. 

Then within linear 
theory the evolution is given by 
\be 
\ddot\delta+2H\dot\delta-\nabla^2\phi=\ddot\delta+2H\dot\delta-\frac{3}{2} 
G_N\fcl H^2\om(a)\delta=0 \ , 
\ee 
where $\om(a)=8\pi\rho_m(a)/(3H^2)$.  The term involving 
the first derivative is called the friction term, and the term 
without derivatives of the perturbation is the source term. 

To make the role of the expansion history more explicit, it is useful 
to write the equation as (e.g.\ \cite{linjen}) 
\bea 
g''&+&\left[4+\frac{1}{2}(\ln H^2)'\right]\,g' \notag\\ 
&+&\left[3+\frac{1}{2}(\ln H^2)'-\frac{3}{2}\fcl(a)\,\om(a)\right]\,g=0\ , 
\label{eq:ddg} 
\eea 
where $g=\delta/a$ and a prime denotes a derivative with respect to 
$\ln a$.  Note that $(\ln H^2)'=-2(1+q)=-3(1+w_{\rm tot})$, where 
$q$ is the deceleration parameter and $w_{\rm tot}$ is the total equation 
of state, or pressure to energy density, ratio.  We will also be 
interested in the logarithmic growth rate $f=1+(\ln g)'$, but note the 
differential equation for $f$ is not linear. 

In the standard case, Eq.~(\ref{eq:ddg}) depends wholly on the expansion 
history, 
through $H(a)$ directly and $\om(a)$.  However, in the general derivation 
from the equations of motion other physical ingredients can enter: 
the equation of state or sound speed of the clustering component, 
deviations from general relativity, or additional forces (e.g.\ early time 
baryon-photon coupling).  See, e.g., \cite{kodamasasaki,mabert,hwang} for 
the general relativistic expressions for a general component.  Again, 
we take the matter component to be pressureless, as usual. 

The growth equation is quite different if we couple the dark matter 
to the dark energy, say.  In this case we would write for the density 
evolutions 
\bea 
\rho_m'&=&-3\rho_m(1+w_m)+Q(a,\phi,\phi') \label{eq:rhomQ} \\ 
\rho_\phi'&=&-3\rho_\phi(1+w_\phi)-Q(a,\phi,\phi') \ , 
\eea 
where the dark energy component involves a scalar field $\phi$, $Q$ 
is the interaction term, and $w$ the equation of state parameter of a 
component.  The equal and opposite interaction terms 
guarantee that total energy conservation holds.  Note that even when 
the dark matter is pressureless, $w_m=0$, there is still an effective 
pressure and hence nonzero effective matter equation of state.  

In the interacting case we must specify several quantities; in addition 
to deciding whether $w_m$ is standard (zero) or not, we must specify 
$w_\phi$ or the dark energy potential (assuming it is a canonical scalar 
field), and the function $Q$.  In the simplest (and probably best 
motivated) case, a Yukawa coupling (linear in $\phi$) in the action 
induces $Q$ linearly proportional to the product of $\phi'$ and the 
trace of the matter energy momentum tensor (e.g.\ $\rho_m(a)$) 
\cite{amen9908023}.  However a considerable range of functions $Q$ have 
been used in the literature.  Moreover, Eq.~(\ref{eq:rhomQ}) indicates 
that matter is being created (or destroyed) -- we must specify how this 
is happening, e.g.\ uniformly in the CMB frame, the dark matter rest frame, 
or some other way (see, e.g., \cite{valiviita}), i.e.\ what is the velocity 
field? 

The interaction gives rise to several effects in the matter density 
perturbation evolution equation: 1) it changes the expansion history $H(a)$, 
and hence the source and friction terms, but also 2) changes the Poisson 
equation or effective gravitational attraction due to the extra fifth force 
of the interaction, which alters the source term, and 3) modifies the 
velocity field evolution in the Euler equation, affecting the friction 
term.  In the source term there is an enhanced attraction (since scalars 
give an attractive force) and diluted potential [since either destruction 
of (clustered) matter or homogeneous (unclustered) creation of matter 
dilutes the potential].  The friction term depends on the form of 
$Q$, but generally creation of matter out of the dark matter rest frame 
will induce an extra drag, slowing growth. 

Due to the myriad uncertainties concerning coupled dark matter, we do not 
consider it further.  Rather we investigate the straightforward model of 
dark matter with self interaction, or general clustering fraction or strength. 
Here the only effect is through the source term.  
Before proceeding to the numerical solutions, we 
explore the analytic growth parameter formalism to help guide our 
physical intuition.

%%%%%%%%%%%%%%%%%%%%%%%%%%%%%%%%%%%%%%%%%%%%%%%%%%%%%%%%%%%%%%%%% 
\section{Comparison to Analytic Growth Parameter Formalism} \label{sec:anly} 

The growth rate $f$ in the standard clustering case has a quadrature 
solution given by \cite{lincahn}.  
We rederive this without assuming $\fcl=1$.  
In the limit that the dark energy density contribution $\Omega_\phi(a)$ 
is small, and hence $f$ deviates little from unity, the solution simplifies 
to 
\be 
f=[a^4 H(a)]^{-1}\int_0^a \frac{dA}{A} \,A^4 H(A) 
\left[1+\frac{3}{2}\fcl(A)\,\om(A)\right] \,. 
\ee 

We now evaluate this for an expansion history from a (generic clustering) 
matter component and a dark energy component $\Omega_\phi$.  Remaining within 
the matter dominated limit, one finds 
\bea 
f&=&a^{-5/2}\int_0^a dA\,A^{3/2}\left[1+\frac{3}{2}\fcl+\frac{1}{2}\left(1-\frac{3}{2}\fcl\right)\op(A)\right] \notag\\ 
&&-\frac{1}{2}a^{-5/2}\op(a)\int_0^a dA\,A^{3/2} 
\left[1+\frac{3}{2}\fcl\right] \label{eq:finteg}\\ 
&\to&\frac{2[1+(3/2)\fcl]}{5}-\frac{1+(3/2)\fcl}{5}\op(a) \notag\\ 
&&+\frac{1-(3/2)\fcl}{2}\, a^{-5/2}\int_0^a dA\,A^{3/2}\op(A) \ , \label{eq:fop} 
\eea 
where the arrow denotes the case where $\fcl(a)$ is constant.  The first 
term will dominate during matter domination, when $\op(a)\ll1$, 
and we recover $f=1$ for $\fcl=1$. 

Recall that this quadrature ignored $(f-1)^2$ terms, and so is valid in the 
limit $\fcl\approx1$.  However, 
we can derive an exact expression for the constant term as asymptotically 
$\op(a)\to0$, i.e.\ pure matter domination, through examining the 
characteristic equation of Eq.~(\ref{eq:ddg}) (using $g$ rather than $f$ 
evades the nonlinearity of the $f$ differential equation).  The result is 
\be 
\finf=\frac{-1+\sqrt{25-24(1-\fcl)}}{4} \ , \label{eq:finf} 
\ee 
which indeed agrees with $(2/5)[1+(3/2)\fcl]$ in the limit of small $\fcl-1$.  
(See \cite{fry,lin0801} for this expression arising in other models.)  
Note that decreasing $\fcl$ decreases $f$, suppressing growth as expected 
from the decreased contribution to the source term clustering density (or 
gravitational potential).  

The dark matter clustering strength could evolve with time.  In this case, 
in the limit of pure matter domination and taking 
$\fcl=\fcli+(\fclt-\fcli)a^s$, with $s<-3w$ so that we see the effect from 
the nonstandard clustering rather than the usual (negligible) early time 
dark energy, 
\be 
f\approx 1-\frac{3}{5}(1-\fcli)+\frac{3}{5+2s}(\fclt-\fcli)\,a^s \ . 
\label{eq:fas} 
\ee 
For example, we might want to choose $\fcli=1$ to suppress early time 
deviation in the growth.  Or such a choice could reflect that the epoch of 
appreciable dark matter self interaction occurs late.  
For simplicity, though, until Sec.~\ref{sec:vary} we will take $\fcl$ to 
be constant.

%%%%%%%%%%%%%%%%%%%%%%%%%%%%%%%%% 
\subsection{Growth index $\gamma$} \label{sec:gamma} 

A simple analytic formula for the density perturbation evolution, valid 
for a wide range of cosmologies including various dark energy models and 
deviations from general relativity (we discuss some exceptions in the next 
subsections), was given by \cite{lingam} as 
\be 
g(a)=e^{\int_0^a (dA/A)\,[\om(A)^\gamma-1]} \ , \label{eq:ggamma} 
\ee 
where $\gamma$ is the growth index.  In terms of the growth rate, 
one can derive $\gamma$ as 
\be 
\gamma=\frac{\ln f}{\ln\om(a)}\approx\frac{1-f}{\op(a)} \ , \label{eq:gammadef} 
\ee 
and remarkably $\gamma$ is effectively constant, despite being the ratio 
of two time dependent quantities.  A constant $\gamma$ reproduces the exact 
numerical solution for the growth evolution for an array of 
cosmologies to better than 0.2\% \cite{lingam,lincahn}.  

If we apply Eq.~(\ref{eq:gammadef}) to the dark matter clustering strength 
case, however, we see from Eq.~(\ref{eq:fop}) that the first, constant term 
in $1-f$ leads to a blow up as we divide by a small, early time $\op(a)$.  
This is because we never have a standard $f=1$ early asymptote in the 
case where $\fcl\ne1$ at early times.  Essentially we never have the 
standard initial condition of $\delta\propto a$, but rather 
\be 
\delta \propto a^{\finf} \ , 
\ee 
where $\finf$ is given by Eq.~(\ref{eq:finf}).  This is not a unique 
situation, and the solution to the problem is already known, as discussed 
in the next subsection.

%%%%%%%%%%%%%%%%%%%%%%%%%%%%%%%%%%%%% 
\subsection{Growth calibration $g_\star$} \label{sec:gstar} 

While the growth index $\gamma$ gives a highly accurate characterization 
of growth under many circumstances, its derivation relies on a period of 
standard matter domination.  If physics is introduced that changes this, 
so that the initial conditions no longer lead to $\delta\propto a$, then 
we must recalibrate the initial growth.  This can be done through the 
growth calibration parameter $\gstar$ \cite{gstar}, shown to be accurate 
to 0.2\% for cases such as early dark energy density, an early period of 
acceleration, some early time changes to the gravitational coupling 
(Newton's constant), or weak coupling between dark matter and dark energy. 

The expression for the growth factor then has a simple modification from 
Eq.~(\ref{eq:ggamma}) to 
\be 
g(a)=\gstar e^{\int_0^a (dA/A)\,[\om(A)^\gamma-1]} \ . \label{eq:ggstar} 
\ee 
Note that this form preserves the growth index $\gamma$ as it was, i.e.\ 
$\gamma$ characterizes the late time growth while $\gstar$ takes care of the 
deviation from the nonstandard early time behavior.  See \cite{gstar} for 
more details about the accuracy and analytic virtues of this form. 

For dark matter with clustering strength $\fcl$ we can 
derive an approximation 
\be 
\gstar\approx 1-4.8(1-\fcli) \ . 
\ee 
Checking numerically by using the exact solution on the left hand side 
of Eq.~(\ref{eq:ggstar}), we find that a constant calibration factor 
$\gstar$ and the standard $\gamma$ gives an excellent fit for the 
integrated growth.  The 
overall accuracy for the growth factor approximation Eq.~(\ref{eq:ggstar}) 
relative to the exact numerical solution is $\sim0.2\%|(1-\fcl)/0.01|$ 
over the range $z=0.25$--2, where most growth observations would be, or 
roughly a factor of two worse over $z=0$--3.  

However, unlike most of the cases discussed in \cite{gstar}, in the 
clustering model the modification to the matter source term does not occur 
only at early times, but rather persists.  This means that while 
the growth calibration may accurately capture the main effects, the 
instantaneous growth rate $f$ is less well approximated.  Recalling that 
$f=d\ln g/d\ln a+1$, we see that a constant $\gstar$ has no effect on 
$f$.  Then $f$ would be biased by $\sim 0.6\%[(\fcl-1)/0.01]$.  
Thus to account for the persistent effect on 
the growth rate while keeping $\gamma$ constant, we would need to make 
$\gstar$ a function of time.  Otherwise, quantities involving $f$ such 
as $fg$, linearly proportional to the quantity $f\sigma_8(z)$ observable 
through redshift space distortions, would be inaccurate.  
Fortunately, we can do better.

%%%%%%%%%%%%%%%%%%%%%%%%%%%%%%%%%% 
\subsection{Growth rate calibration $\finf$} \label{sec:finf} 

While the growth rate $f$ in the case of nonstandard matter clustering 
evolution 
does not accord perfectly with the constant growth index formalism, we 
find that the ratios of the growth rates at different redshifts do.  
That is, rather than calibrate the growth factor through the early time 
$\gstar$ factor, it is more effective to calibrate the growth rate 
through the factor $\finf$ already introduced in Eq.~(\ref{eq:finf}).  
This not only accounts for the early time initial conditions, but the 
persistence of the altered growth rate at later times.  

The rate calibrated version of the analytic growth parameter formalism 
becomes 
\be 
g(a)= e^{\int_0^a (dA/A)\,[\finf\om(A)^\gamma-1]} \ . \label{eq:gfinf} 
\ee 
Comparing to the exact numerical solution we find that $\gamma$ 
remains virtually unchanged from the standard case, as it should within 
its interpretation as a gravitational growth index (since we have not 
changed the theory of gravity).  Since 
$\finf$ is determined by Eq.~(\ref{eq:finf}), then this fit has no 
free parameters.  Remarkably, its accuracy is $0.25\%|(1-\fcl)/0.01|$ 
in growth at all redshifts, and does not degrade for quantities involving 
the growth rate, such as the observational quantity $f\sigma_8(z)$, 
for $z>0.25$.  

Indeed the growth rate itself 
\be 
f=\finf\,\om(a)^\gamma 
\ee 
is accurate to 0.2\% even for $\fcl=0.9$ for $z>0.25$ (and to 0.1\% for 
$z>0.4$), as is the quantity 
$f\sigma_8(z)/\sigma_{8,0}$ discussed in the next section. 
This multiplicative form for the growth rate 
was also shown to work for some coupled dark matter-dark energy models 
by \cite{ameneta}, to $\sim$1\% in the integrated growth, where their $\eta$ 
corresponds to our $\finf-1$ and includes a fit parameter. 

\begin{figure}[htbp!]
\includegraphics[width=\columnwidth]{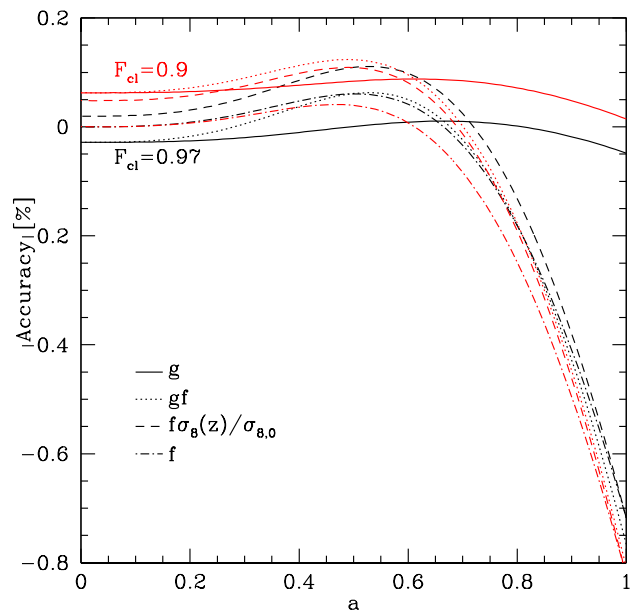} 
\caption{The fitting form of Eq.~(\ref{eq:gfinfs}) predicting the 
influence of dark matter clustering strength $\fcl$ on growth variables is 
accurate to better than 0.1\% relative to numerical solutions, for $z>0.4$. 
The growth factor itself is fit to better than 0.1\% for all redshifts. 
} 
\label{fig:acc} 
\end{figure}

Since the clustering strength affects the initial conditions on both 
the growth factor and growth rate, and 
the late time growth, we can incorporate both the growth calibration 
$\gstar$ and the growth rate calibration $f_\infty$, if we want to achieve 
even better accuracy.  We find that when calibrating the growth rate 
according to Eq.~(\ref{eq:finf}), the remaining growth calibration is 
given by 
\be 
\tilde g_\star\approx 1+0.255(1-\fcl) \ . 
\ee 
The final version of the analytic growth formalism becomes 
\be 
g(a)= \tilde g_\star e^{\int_0^a (dA/A)\,[\finf\om(A)^\gamma-1]} \ . 
\label{eq:gfinfs} 
\ee 
We emphasize that the single parameter $\fcl$ determines both 
$\tilde g_\star$ and $\finf$. 
The resulting growth factor, growth rate, and observable combinations 
agree with numerical solutions of the growth differential equation to 
0.1\% for $z>0.4$, even for extreme models with $\fcl=0.9$, as 
shown in Fig.~\ref{fig:acc}.

%%%%%%%%%%%% 

We now turn to observational measures of the growth rate from redshift 
space distortions, and will use the exact numerical solutions for 
comparison.

%%%%%%%%%%%%%%%%%%%%%%%%%%%%%%%%%%%%%%%%%%%%%%%%%%%%%%%%%%%%%%%%% 
\section{Growth Rate Evolution and Measurements} \label{sec:data} 

\subsection{Observables} \label{sec:obs}

The galaxy power spectrum in real space is isotropic, but when observed 
in redshift space it gains anisotropic contributions due to the galaxy 
peculiar velocities.  These are referred to as redshift space distortions 
and are proportional to the growth rate $f$ (see \cite{hamilton} for a 
review).  In the linear limit the Kaiser formula \cite{kaiser} for 
the redshift space power spectrum $P^s$ is 
\be 
P^s(k,\mu,z)=(b+f\mu^2)^2\,P^r(k,z) 
\ee 
where $\mu$ is the cosine of the angle of the density perturbation Fourier 
mode $k$ with respect to the line of sight, and $b$ is the galaxy bias. 
Since the linear theory power spectrum $P^r$ is proportional to the mass 
amplitude fluctuation squared, $\sige^2$, then the three cosmological 
quantities of observational interest are $f\sige$ and $b\sige$ or the 
ratio $f/b$ \cite{percwhite}, all of which are redshift dependent. 

\begin{figure}[htbp!]
\includegraphics[width=\columnwidth]{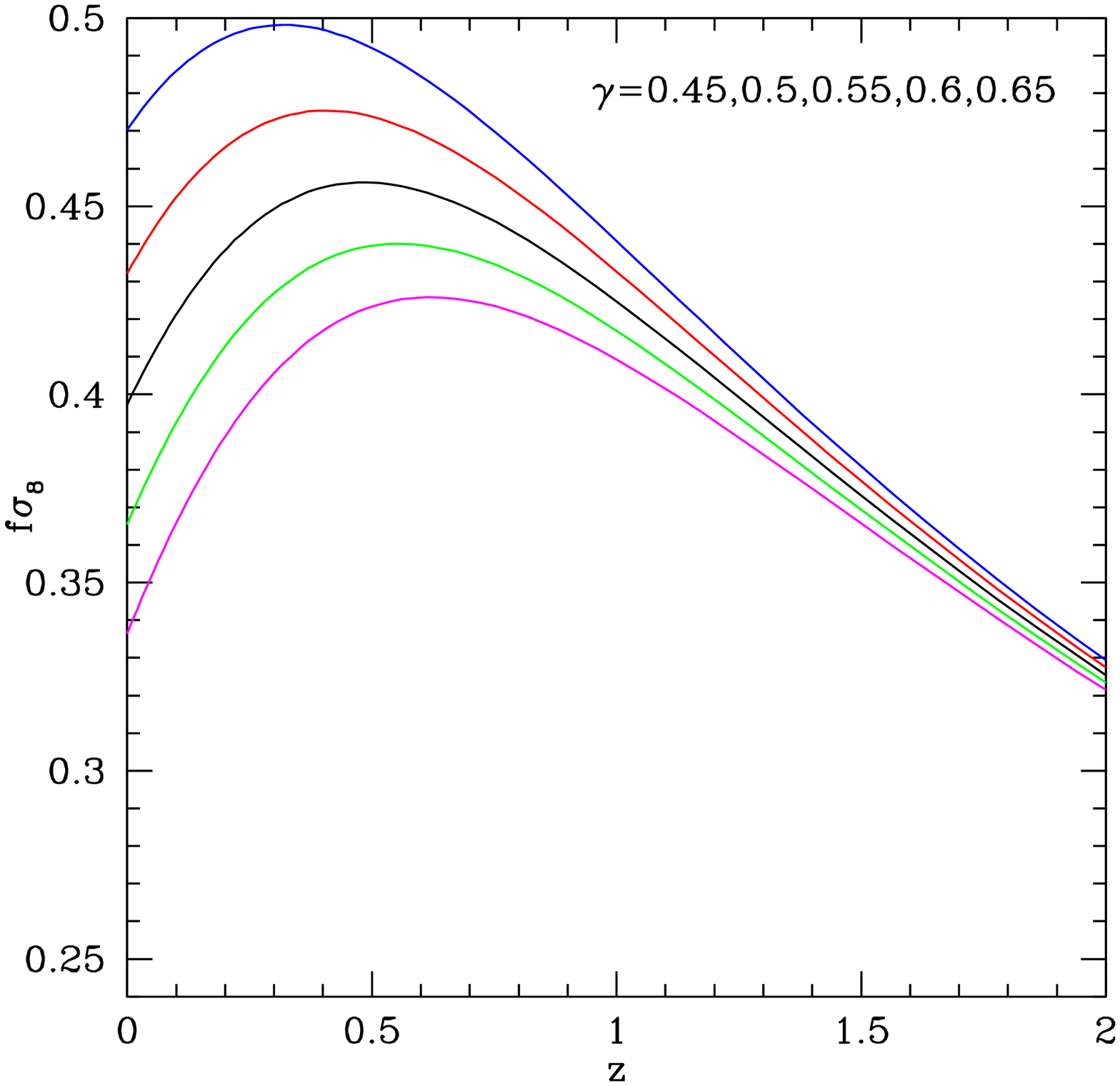}\\ 
\includegraphics[width=\columnwidth]{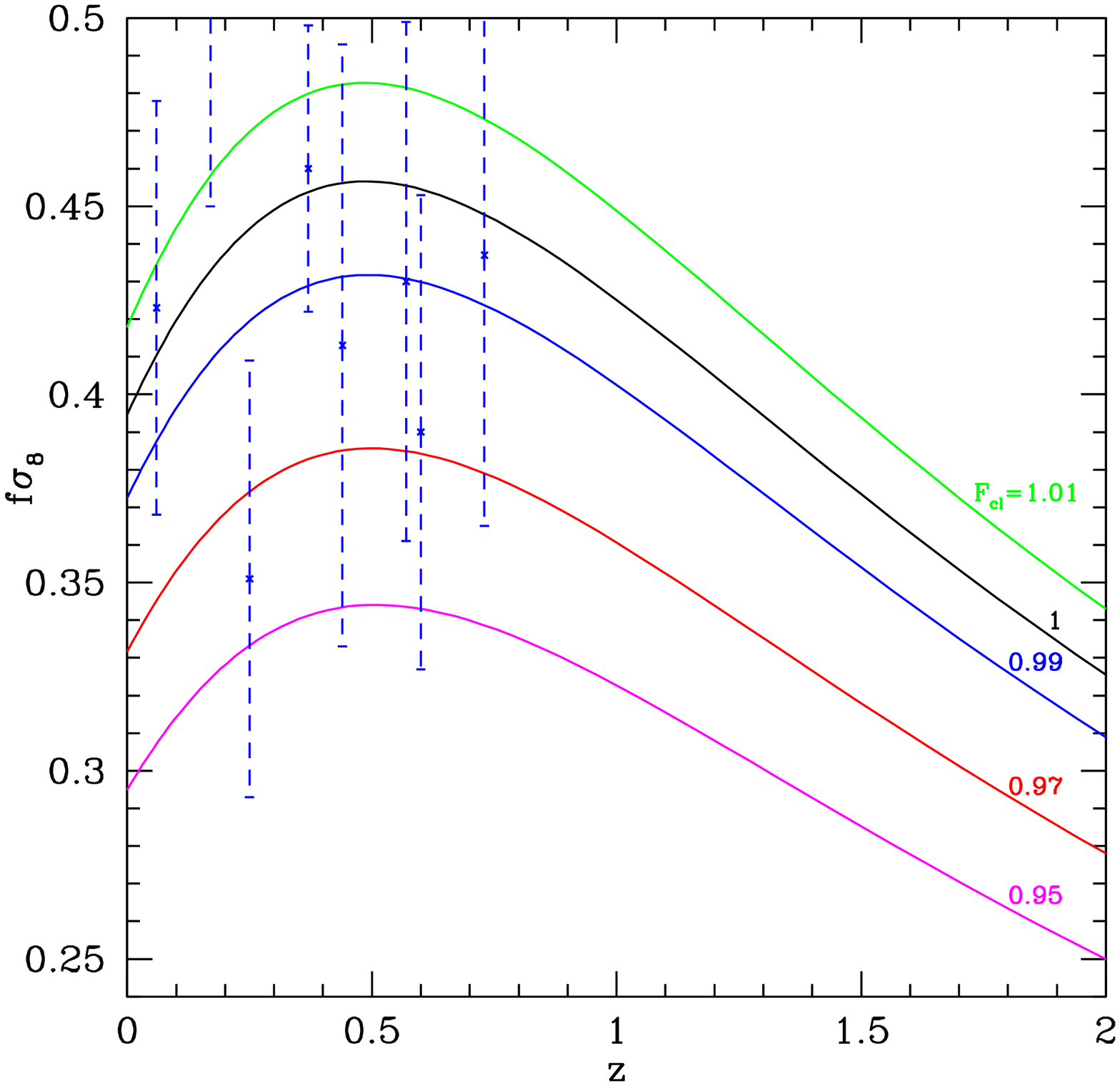} 
\caption{The growth rate $f\sigma_8(z)\propto d\delta/d\ln a$ is plotted 
for different values of the growth index $\gamma$, for standard matter with 
$\fcl=1$ (top panel), and different values of the dark matter clustering 
strength $\fcl$ (bottom panel).  Dashed error bars show measurements from 
several surveys \cite{fsigdata}. 
} 
\label{fig:fsig} 
\end{figure}

Figure~\ref{fig:fsig} plots the behavior of $f\sige(z)$ with the growth 
index $\gamma$ and dark matter clustering strength $\fcl$.  
We vary $\gamma$, i.e.\ do not assume general relativity, in this one 
place in order to contrast the effects of $\gamma$ and $\fcl$.  
The background expansion is taken to be 
$\Lambda$CDM and the primordial scalar amplitude $A_s$ is fixed.  
The different 
$\gamma$ values (taking the standard case of $\fcl=1$) give rise to 
$f\sige$ differing most at low redshift, since at high redshift all go 
to standard matter domination.  Stronger gravity (smaller $\gamma$) 
increases the growth rate and shifts the growth suppression due to dark 
energy to later times, so the small $\gamma$ curves have higher peaks, 
at lower redshift.  For different $\fcl$, however, smaller $\fcl$ 
acts to suppress the growth rate, and this persists to high redshift, 
so the curves remain distinct.

%%%%%%%%%%%% 

\begin{figure}[htbp!]
\includegraphics[width=\columnwidth]{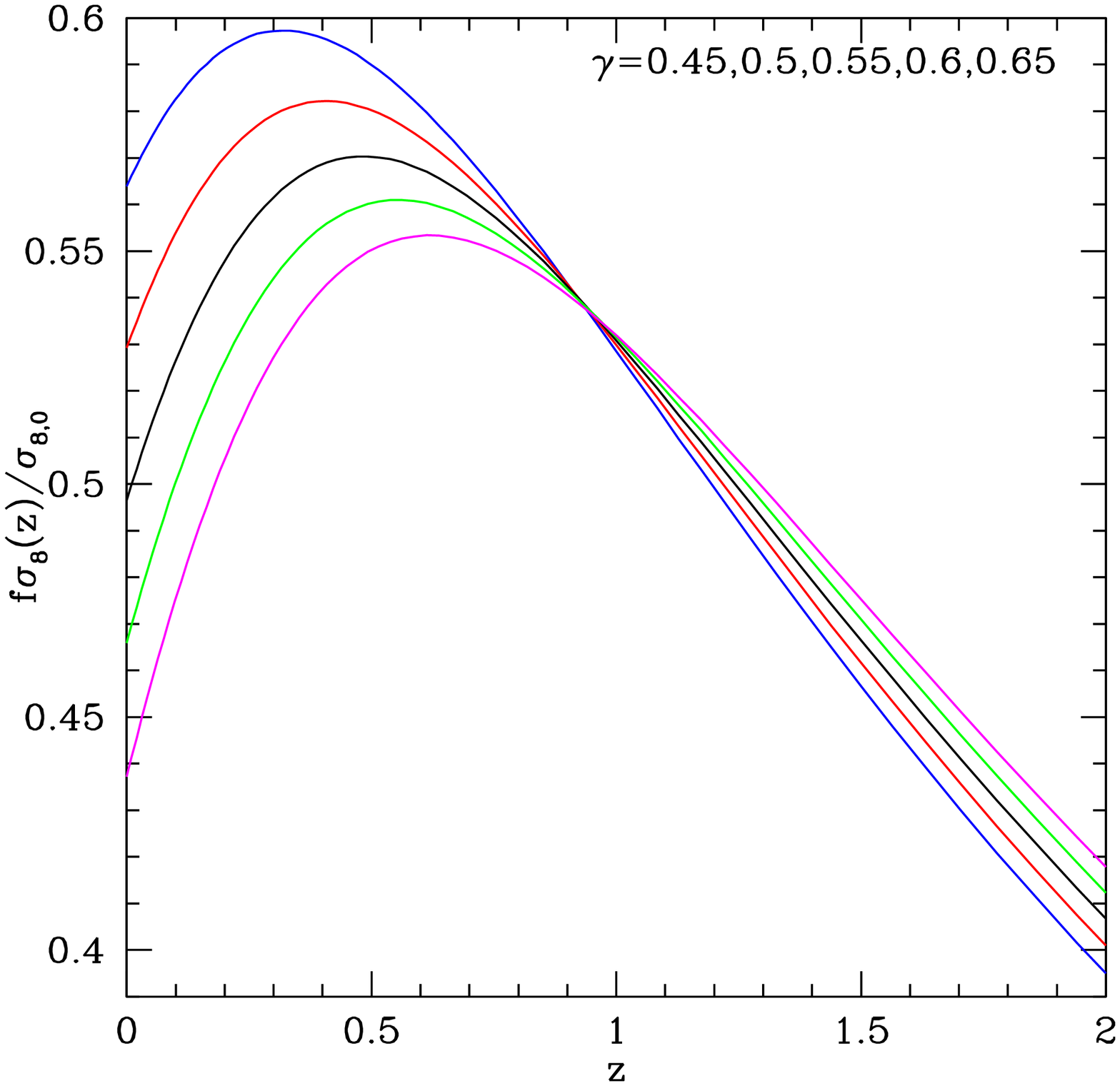}\\ 
\includegraphics[width=\columnwidth]{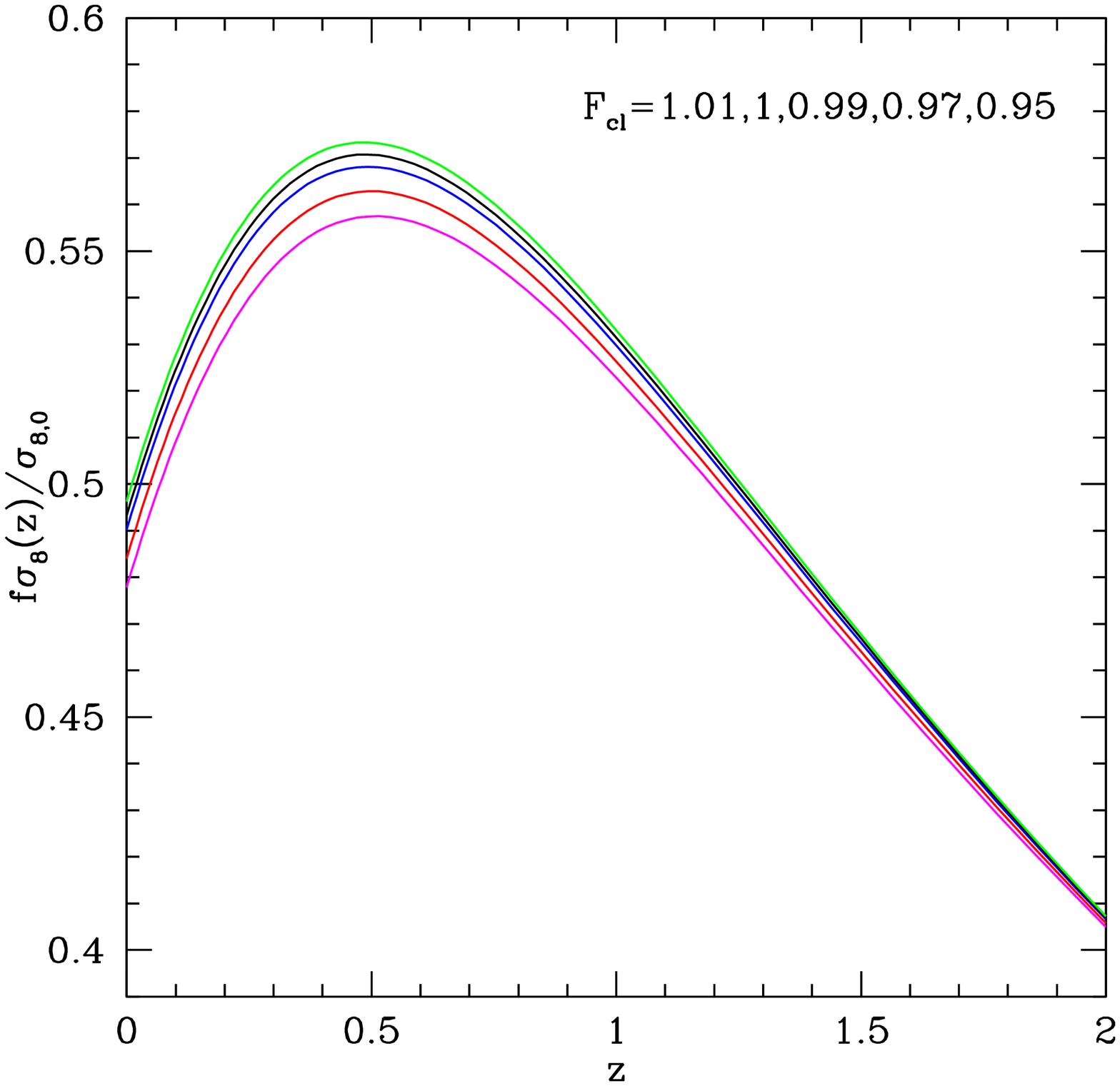} 
\caption{As Fig.~\ref{fig:fsig}, for the distortion factor 
$f\sige(z)/\siget\propto f/b$. 
} 
\label{fig:fsig0} 
\end{figure}

Galaxy bias is an astrophysical quantity, but observations indicate a 
good fit for the types of galaxy targeted in galaxy redshift surveys is 
given by \cite{padma} 
\be 
b(z)=b_0 \frac{\delta(z=0)}{\delta(z)} \ , 
\ee 
so that $b\sige$ is nearly constant.  In this case, the observational quantity 
$\beta(z)\equiv f/b\propto f\sige/\siget$.  Figure~\ref{fig:fsig0} displays 
this for the various values of $\gamma$ and $\fcl$.  An interesting null 
occurs where the $\gamma$ dependence vanishes at $z\approx0.95$.  For the 
$\fcl$ dependence, the differing total growth factors $\siget$ almost cancel 
out the growth rate differences, leading to little distinction at high 
redshift.

%%%%%%%%%%%% 

Comparing current redshift space distortion data for $f\sige$ from the 
several surveys \cite{fsigdata}, we find that the data are consistent 
with $\fcl\approx0.99\pm0.02$, holding all other parameters 
fixed.  For more realistic constraints we must take into account parameter 
covariances, while keeping in mind that future observations will have 
better precision for constraining the behavior of dark matter.  We should 
also consider the clustering strength to vary with time if we want to 
preserve the standard early universe.

%%%%%%%%%%%%%%%%%%%%%%%%%%%%%%%%%%%%%%% 
\subsection{Time varying clustering} \label{sec:vary} 

For a constant deviation in dark matter clustering strength, the impact 
on observables builds up over cosmic history.  But as mentioned previously, 
we generally want to avoid early time deviations to preserve the CMB power 
spectrum agreement with the standard dark matter scenario.  For example, 
gravitational potentials would decay roughly as 
$\phi\sim a^{-3(1-\fcli)/5}$, leading to a large integrated Sachs-Wolfe 
(ISW) effect.  Therefore a more viable model would have $\fcli=1$, but 
we still want to explore deviations at late times (where dark energy breaks 
matter domination and gives an ISW effect anyway).  

The time variation of the clustering strength may be driven by the matter 
density, and hence the expansion factor $a$, as in the model used in 
Eq.~(\ref{eq:fas}), or have a time scale from dark matter properties as 
in dark matter decay models, or be somehow driven by the cosmic acceleration. 
Without a physical model we resort to phenomenology, and explore the second 
two models, calling them the step and $\Omega_\phi$ models: 
\bea 
\fcl(a)&=&\begin{cases} \fcli &z\ge1\\ \fclt &z<1 \ . \end{cases} \label{eq:stepmodel}\\ 
\fcl(a)&=&\fcli+(\fclt-\fcli)\frac{\Omega_\phi(a)}{\Omega_{\phi,0}} \label{eq:owmodel} 
\eea 

\begin{figure}[htbp!]
\includegraphics[width=\columnwidth]{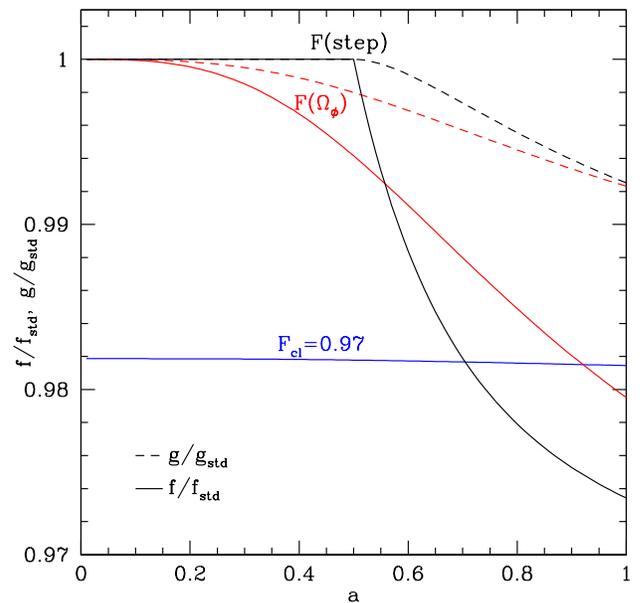} 
\caption{Time varying dark matter clustering strength allows preservation 
of early universe behavior and the CMB, while giving late time signatures. 
The solid curves shows the growth rate and the dashed curves the growth 
factor, each relative to the standard $\fcl=1$ case, for 
the step transition model of Eq.~(\ref{eq:stepmodel}), the $\Omega_\phi$ 
model of Eq.~(\ref{eq:owmodel}), and constant clustering strength 
$\fcl=0.97$ (only $f/f_{\rm std}$ is shown here since $g/g_{\rm std}$ is 
way down at 0.86).  
The former two models take $\fcli=1$, $\fclt=0.94$.  
} 
\label{fig:trans} 
\end{figure}

Figure~\ref{fig:trans} illustrates the behaviors of these models in 
comparison with the standard clustering $\fcl=1$ model.  
Note that these time 
varying models are much more similar in growth factor to the standard 
clustering case, while retaining a distinction from it in 
the growth rate.  For simplicity we will adopt for the remainder of the 
article the $\Omega_\phi$ model with $\fcli=1$, thus keeping dark matter 
clustering at CMB last scattering unchanged and the growth factor (and ISW) 
comparable to the standard cosmology.

%%%%%%%%%%%% 

%%%%%%%%%%%%%%%%%%%%%%%%%%%%%%%%%%%%%%%%%%
\subsection{Projected constraints} 

From the shape of the curves in Fig.~\ref{fig:fsig} we expect the 
redshift space distortion 
observable $f\sige(z)$ to have covariance with the present amplitude of 
mass fluctuations, $\siget$.  In addition, time varying dark energy 
density will also change matter perturbation growth and so fitting for 
the expansion history simultaneously with the growth could open up 
degeneracies.  

To attempt precise constraints on the clustering strength 
we consider future data sets comprised of growth rate measurements of 
$f\sige$ to 1\% at $z=0.1, 0.3,\dots 1.5$, together with measurement of 
the distance to CMB last scattering to 0.2\% and distance data from 
1800 supernovae from $z=0$--1.5 with magnitude systematic floor of 
$0.02(1+z)/2.5$. 
The last two data sets help constrain the matter density $\om$ and the 
dark energy equation of state parameters $w_0$, $w_a$ where 
$w(z)=w_0+w_a z/(1+z)$.  We carry out a Fisher matrix analysis over the 
parameter set $\{\om,w_0,w_a,\siget,\fclt,{\mathcal M}\}$, where 
${\mathcal M}$ is a nuisance parameter for the supernova magnitude-distance 
calibration.  For the dark matter clustering strength we use the 
$\Omega_\phi$ model of Eq.~(\ref{eq:owmodel}), which has a single, new free 
parameter $\fclt$, the clustering strength today, fixing the early time 
clustering strength $\fcli=1$. 

The previous intuition is borne out, with $\fclt$ having correlation 
coefficients of $-0.93$, $-0.96$, $0.78$ with $\siget$, $w_a$, $w_0$. 
The uncertainty on the clustering strength in this time varying model, 
marginalized over all the other parameters, is $\sigma(\fclt)=0.28$, even 
with the optimistic measurement precisions used; contrast this with the 
unmarginalized uncertainty of 0.026.  In order to obtain more robust 
constraints on dark matter clustering strength, one must abandon 
simultaneous fitting for growth and expansion.  If one assumes that the 
background expansion is $\Lambda$CDM, so deviations from this form for the 
growth arise solely from the clustering strength, then the constraints 
tighten substantially.  Marginalizing over $\{\om,\siget,\fclt,{\mathcal M}\}$ 
then delivers $\sigma(\fclt)=0.042$. 

\begin{figure}
\includegraphics[width=\columnwidth]{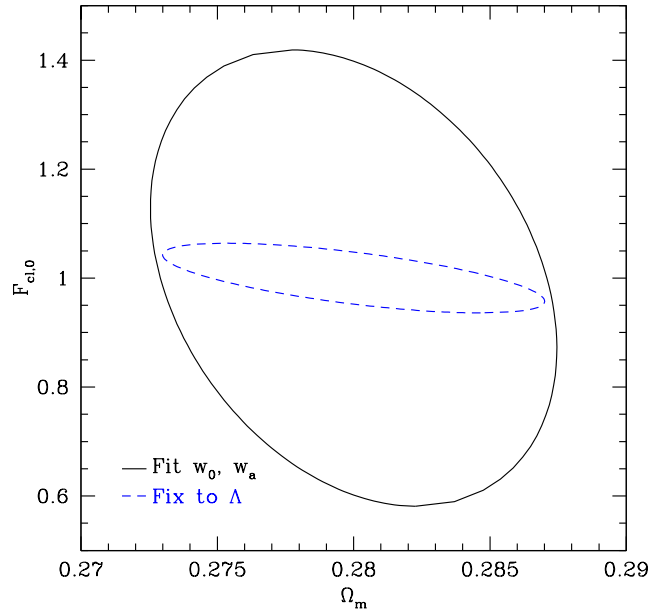} 
\caption{68\% CL contour of the matter density and dark matter clustering 
strength in the growth history is plotted for the case of simultaneous 
fitting for the background expansion (solid, black curve) and fixing it 
to $\Lambda$CDM (dashed, blue curve).  Redshift space distortions, in 
conjunction with CMB and 
supernova (or baryon acoustic oscillation) distances, can only reasonably 
constrain dark matter clustering strength if the background expansion is 
fixed. 
} 
\label{fig:omf0} 
\end{figure}

Figure~\ref{fig:omf0} shows joint constraints on the matter density and 
matter clustering strength, marginalizing over the other parameters.  One 
clearly sees the effect of fitting for, vs.\ fixing, the non-$\Lambda$CDM 
expansion (i.e.\ $w_0$, $w_a$).

%%%%%%%%%%%% 
%\begin{figure}[htbp!]

%%%%%%%%%%%%%%%%%%%%%%%%%%%%%%%%%%%%%%%%%%%%%%%%%%%%%%%%%%%%%%%%% 
\section{Conclusions} \label{sec:concl} 

Dark matter remains a mysterious component of our universe, without 
identification or clear knowledge of its properties.  Beyond standard cold 
dark matter, one could explore its interactions with other components 
(e.g.\ dark energy) or an equation of state differing from zero pressure. 
Here we concentrate on its clustering strength, whether it interacts 
with gravitational strength or has some anomalous self interaction, or 
whether some fraction of it does not cluster at all. 

Our approach is meant as purely phenomenological, but models exist in the 
literature with interesting, related properties, as mentioned in the 
Introduction.  The growth factor and growth rate calibration approach 
may have broad physical applicability, in the same way the gravitational 
growth index $\gamma$ does. 

For the clustering strength on cosmic scales, growth rate 
measurements through redshift space distortions in galaxy surveys provide 
a probe for deviations from the standard scenario.  Redshift space 
distortions are already used to probe matter density, dark energy, and 
gravity; here we explore their use for probing dark matter clustering 
strength.  

We derive highly accurate analytic fits to the evolution of the growth 
factor and growth rate for constant clustering strength, showing how it 
combines aspects of early time growth calibration and a multiplicative 
change to the growth rate.  The clustering strength remains distinct from 
deviations to gravity appearing in the gravitational growth index $\gamma$. 
Considering only variations due to a {\it time independent\/} 
strength $\fcl$, current 
measurements of $f\sige(z)$ are consistent with a value $\fcl=0.99\pm0.02$, 
where the standard value is $\fcl=1$. 

To avoid deviations in the early universe and CMB, the clustering strength 
should take the standard value at early times, but may transition to a 
different late time value.  This {\it time dependence\/}, 
together with covariance 
with present mass fluctuation amplitude $\siget$ and time dependence of 
dark energy $w_0$, $w_a$, degrades estimation of the clustering strength, 
even with next generation galaxy redshift surveys, yielding 
$\sigma(\fclt)\approx 0.28$.  Only when the background expansion is fixed, 
e.g.\ to $\Lambda$CDM, separating out expansion effects on the growth, 
can growth rate measurements provide strong constraints, 
$\sigma(\fclt)\approx 0.04$.  
While redshift space distortions can test dark matter clustering, 
stringent constraints will require the addition of astrophysical probes 
on galactic and cluster scales.

%%%%%%%%%%%%%%%%%%%%%%%%%%%% 
\acknowledgments 

This work has been supported by DOE grant DE-SC-0007867 and the Director, 
Office of Science, Office of High Energy Physics, 
of the U.S.\ Department of Energy under Contract No.\ DE-AC02-05CH11231, 
and World Class University grant R32-2009-000-10130-0 through the 
National Research Foundation, Ministry of Education, Science and Technology 
of Korea.

%%%%%%%%%%%%%%%%%%%%%%%%%%%%%%%%%%%%%%%%%%%% 

\end{document}